\documentclass[prl,showpacs,twocolumn,floatfix]{revtex4}
\usepackage{epsfig}
\usepackage{mathrsfs}
\usepackage{amssymb}
\usepackage{color}
\newcommand{\bcen}{\begin{center}}
\newcommand{\ecen}{\end{center}}
\newcommand{\btab}{\begin{tabular}}
\newcommand{\etab}{\end{tabular}}
\newcommand{\bdes}{\begin{description}}
\newcommand{\edes}{\end{description}}

\newcommand{\beq}{\begin{equation}}
\newcommand{\eeq}{\end{equation}}
\newcommand{\bea}{\begin{eqnarray}}
\newcommand{\eea}{\end{eqnarray}}

\newcommand{\etal}{{\it et al}}

\newcommand{\bary}{\begin{array}}
\newcommand{\eary}{\end{array}}
\newcommand{\benum}{\begin{enumerate}}
\newcommand{\eenum}{\end{enumerate}}
\newcommand{\bitem}{\begin{itemize}}
\newcommand{\eitem}{\end{itemize}}
\newcommand{\prn}[1] {(\ref{#1})}

\newcommand{\fig}[1]{FIG.~\ref{#1}}
\newcommand{\Fig}[1]{FIG.~\ref{#1}}
\newcommand{\mean}[1]{\langle #1 \rangle}

\newcommand{\Ejt}{E_{{JT}}}

\begin{document}

\title{Coulomb Interactions and Nanoscale Electronic Inhomogeneities in Manganites}

\author{Vijay~B.~Shenoy$^{1,2}$}%\email[]{shenoy@mrc.iisc.ernet.in}
\author{Tribikram Gupta$^2$}%\email[]{gupta@physics.iisc.ernet.in}
\author{H.~R.~Krishnamurthy$^{2,4}$}%\email[]{hrkrish@physics.iisc.ernet.in}
\author{T.~V.~Ramakrishnan$^{2,3,4}$}%\email[]{tvrama@bhu.ac.in}
\affiliation{$^1$Materials Research Centre, Indian Institute of
Science, Bangalore 560 012, India\\$^2$Centre For Condensed Matter
Theory, Indian Institute of Science, Bangalore 560 012,
India\\$^{3}$Department of Physics, Banaras Hindu University,
Varanasi 221 005, UP, India \\$^{4}$Jawaharlal Nehru Centre for
Advanced Scientific Research, Jakkur, Bangalore 560 064, India}

\date{\today}

\begin{abstract}
We address the issue of endemic electronic inhomogeneities in
manganites using extensive simulations on a new model with Coulomb
interactions amongst two electronic fluids, one localized
(polaronic), the other extended (band-like), and dopant ions. The
long range Coulomb interactions frustrate phase separation induced
by the strong on site repulsion between the fluids. A single quantum
phase ensues which is intrinsically and strongly inhomogeneous at a
nano-scale, but homogeneous on meso-scales, with many
characteristics (including colossal responses)that agree with
experiments. This, we argue, is the origin of nanoscale
inhomogeneities in manganites, rather than phase competition and
disorder related effects as often proposed.
\end{abstract}


\maketitle

An intriguing generic characteristic of complex  solids such as
doped manganites\cite{Manganites,Dagotto2003,Sarma2004},
cuprates\cite{Sigmund1994} and cobaltates\cite{Cobaltates} seems to
be the coexistence of patches of metallic and insulating regions
(often dubbed as `phases'). This `electronic inhomogeneity' can vary
from nanometers to microns
%%S  in scale
and can be static or
dynamic\cite{Manganites,Dagotto2003,Sarma2004,Sigmund1994,Cobaltates}.
Questions as to whether this can be characterized as `electronic
softness'\cite{Dagotto2005} and is a {\it defining feature} of these
materials, its origins and role in determining their electronic
properties, etc., are among the most actively explored issues.
%%S in condensed matter physics.
However, proposed
mechanisms\cite{Dagotto2003,Ahn2004} are either at the level of
scenarios or
%%S have been based on
toy models, and do not adequately
address the specifics of the scale and nature of inhomogeneities in
the actual systems.

Motivated by the fact that electronic inhomogeneities came to
prominent attention first in  manganites\cite{Manganites}, in this
letter we address these issues using extensive simulations on a new,
``extended $\ell b$ model''.  This is an extension of the $\ell b$
model, involving two kinds of electronic fluids, one localized and
polaronic ($\ell$), the other extended and band-like ($b$), which
was introduced recently\cite{Pai2003,Ramakrishnan2004} and shown to
describe successfully many puzzling phenomena (including colossal
magnetoresistance) observed in doped manganites of the type
Re$_{1-x}$Ak$_x$MnO$_3$ (Re=rare earth ions such as La, Nd; Ak=
alkaline earth ions, such as Ca, Sr). The extensions invoke {\it
long range Coulomb interactions amongst the two fluids, and dopant
ions}. We explore here their consequences, especially in the context
of issues connected with electronic inhomogeneities where they play
a crucial role.

The $\ell b$ model\cite{Pai2003,Ramakrishnan2004} is based on the
idea that under the conditions prevailing in the doped manganites,
the electrons populating the doubly degenerate $e_g$ states centered
at the Mn sites spontaneously reorganize themselves into two types
of electron fluids which coexist. One is obtained by populating
essentially site localized states labeled $\ell$ which are
polaronic, with strong local Jahn-Teller(JT) distortions of the
oxygen octahedra, an energy gain $\Ejt \sim 0.5$ eV\cite{Material}
and exponentially reduced intersite hopping. The other, labeled $b$,
is a fluid of broad band, non-polaronic electrons, with no
associated lattice distortions, and undiminished hopping amplitudes.
For a generic $x$ in the regime $(0.1 \lesssim x \lesssim 0.4)$ of
interest to us in this paper, manganites do not exhibit orbital long
range order and can be regarded as `orbital liquids'. Hence one can
characterize the hopping of the $b$ electrons by a single orbitally
averaged number $t \, (\sim 0.2)$ eV and ignore the $e_g$ orbital
index.  There is a strong local repulsion between the two fluids, as
double occupancy on a polaronic site costs an extra Coulomb energy
$U \, (\sim 5 eV)$.  The spins of $\ell$ and $b$ are enslaved to the
Mn-$t_{2g}$ spins $(S = 3/2)$ on site due to the large ferromagnetic
Hund's coupling $J_H \, (\sim 2 eV)$. Furthermore, there is a {\it
new, occupancy dependent, ferromagnetic nearest neighbor exchange
coupling}\cite{Ramakrishnan2004} between the $t_{2g}$ core spins, of
order $x(1-x)t^2/\Ejt$, referred to as ``virtual double exchange'',
which overwhelms the normal super-exchange for $x \gtrsim 0.1$.
Hence, as is seen in manganites over the above mentioned range of
doping $x$, the ground state is ferromagnetic (insulating or
metallic). In the simplest picture, assuming all the $t_{2g}$ spins
and the $e_g$ spins to be aligned parallel, the $\ell$ and $b$
electrons can be regarded as spin-less, leading naturally to the
Falicov-Kimball\cite{Freericks2003} like $\ell b$ Hamiltonian
{\small \beq H_{\ell b} = -\Ejt \sum_i n_{\ell i} - t \sum_{\langle
ij \rangle}(b^\dagger_i b_j + \mbox{h.~c.}) + U \sum_i n_{\ell i}
n_{b i} \label{purelb} \eeq} Here ${\ell^\dagger_i}$ and
${b^\dagger_i}$ create $\ell$ polarons and $b$ electrons
respectively at the sites ${i}$ of a cubic Mn lattice, and $n_{\ell
i} \equiv \ell^\dagger_i \ell_i $ and $n_{b i} \equiv b^\dagger_i
b_i $ are the corresponding number operators.

\begin{figure}
\centerline{\epsfxsize=8.0truecm \epsfbox{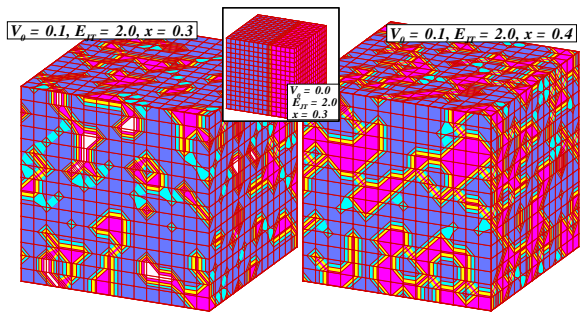}}
\caption{(color online) Real space electronic distribution obtained
from simulations on a 16$^3$ cube. Magenta (darkest)
 denotes hole clumps with occupied $b$ electrons,
white (lightest) denotes hole clumps with no $b$ electrons, cyan
(2nd lightest)  denote singleton holes, and
light blue (2nd darkest)  represents regions with $\ell$ polarons.
The configuration on the left  shows isolated clumps with occupied
$b$-electrons ($b$-electron puddles).  For larger doping,
percolating clumps are obtained and the system is a metal (right).
The inset is results in the {\it absence} of long range Coulomb
interaction ($V_0 = 0.0$) and shows `macroscopic phase separation'.
All energy scales are in the units of $t$. } \label{clump}
\end{figure}

In this Letter we extend the above model to include the long range
Coulomb interactions that are necessarily present in the doped
manganites. This is done at the simplest level by associating {\it
quenched} (spatially fixed) charges $-|e|$ (with respect to the
Re$^{3+}$ background) at a fraction $x$ of random, body centered, `Ak
sites', and {\it annealed} charges $+|e|$ with the corresponding
deficits in the Mn-$e_g$ electron occupancy (overall reduced to
$(1-x)$ per site). In terms of the hole operator ($h^\dagger_i
\equiv \ell_i$ which removes an $\ell$ polaron at site $i$), and the
electron charge operator $q_i \equiv h^\dagger_i h_i - b^\dagger_i
b_i$, which has the average value $x$ per site because of overall
charge neutrality, the extended model Hamiltonian is  {\small \beq H
= H_{\ell b} + H_C \,; \, H_C = \sum_{i} \Phi_i q_i + \frac{V_0}{2}
\sum_{i \ne j} \frac{q_i q_j}{r_{ij}}. \label{ElbHam} \eeq} The
Coulomb term $H_{C}$ has two parts; the charge at site $i$ has
energy $q_i \Phi_i$, where $\Phi_i$ is the electrostatic potential
there due to Ak$^{2+}$ ions, and the interaction between the charges
at site $i$ and $j$ leads to an energy $V_0 \frac{q_i q_j}{r_{ij}}$.

\begin{figure}
\centerline{\epsfxsize=6.5truecm
\epsfbox{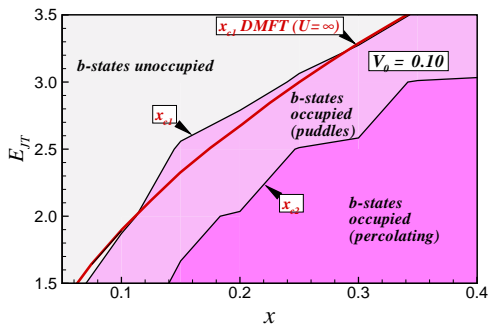} } 
\caption{(color online) Critical doping
levels $x_{c1}$ and $x_{c2}$. $x_{c1}$
separates the lightest region, where no $b$
states are occupied, from the intermediate shade region, which has
$b$ states occupied in  puddles. The darkest shaded
region, with $x > x_{c2}$, has some occupied $b$ states  that
percolate through the 10$^3$ box. The solid line corresponds to
the DMFT result\cite{Pai2003} for $x_{c1}$. } \label{btype}
\end{figure}

We now describe the results of numerical simulations of the above
model \prn{ElbHam} on finite 3d periodic lattices of sizes up
20$^3$. The electrostatic interaction is treated in the Hartree
approximation, using the {\em mean} $b$ electron charge at site $i$
, i.~e., $q_i$ is approximated by $\mean{q_i} = h^\dagger_i h_i -
\mean{b^\dagger_i b_i}$. Furthermore, since $U$ is large, for
simplicity we take the $U \rightarrow \infty$ limit; i.e., assume
that $b$ electrons do not hop to sites occupied by $\ell$, so the
two kinds of electrons form disjoint clusters. On a hole-cluster,
which has two or more $h$-sites each accessible to the other by
electron hops, which we will refer to as a ``clump'', the $b$
electron energy levels are determined exactly by solving the
intersite Hamiltonian $H_b = - t \sum_{ij} (b^\dagger_i b_j +
\mbox{h.~c.})$.  The ground state (i.e, at zero temperature) is
obtained by starting from an overall charge neutral trial
configuration random of $\ell$ polaronic sites,  and performing
electron transfers that lower energy, till none such exist. Some of
these are of the type $\ell-h$ as in a classical Coulomb
glass\cite{CoulombGlass}, in which an $\ell$ electron is moved to a
hole site. In addition there are $\ell-b$ and $b-b$ transfers
involving one or two $b$-states whose energy is quantum mechanically
obtained for a particular clump structure. At each iteration then,
we find the best possible transfer, i.~e., the one which lowers most
the (occupied) single particle levels whose energies include Coulomb
interactions. Then we perform the transfer, and update the clump
structure, if necessary, i.~e., redistribute the $b$ electrons in
the new clumps. The process is repeated till the final
$b$-clump/$\ell$-polaron structure is stable against all further
electron transfers. {\it This is a new generalization of the common
Coulomb glass simulation\cite{CoulombGlass} which includes the
quantum mechanically obtained $b$ states within their clump or
puddle}. The electrostatic energy is calculated accurately using the
Ewald technique and fast Fourier transform routines.

In the discussion below, all length scales are normalized by the
lattice parameter $a$, and we use dimensionless energy parameters
$\Ejt$ and $V_0$, scaled by the hopping amplitude $t$. Realistic
values\cite{Material} for manganites are $2.0 \le \Ejt \le 3.0$ and
$0.01 \le V_0 \le 0.1$ with $t \approx 0.2$eV. In the absence of
$V_0$, the system ``phase separates'', i.~e., holes move to one side
of the box and several $\ell$ polarons are converted to $b$
electrons that occupy the low energy band states with energies below
$-\Ejt$ in this large clump (\Fig{clump}, see inset). This phase
separation {\it is due to strong local correlations} (large $U$
between $\ell$ s and $b$ s) and is in agreement with known results
for the Falicov-Kimball model\cite{Freericks2003}. The presence of
$V_0$ renders this phase separation energetically unfavorable, and
intermixes the phases as expected\cite{Emery}. The favored
electronic configurations strongly depend on the JT energy $\Ejt$
and the doping $x$. Two examples are shown in \Fig{clump}. For a
given $\Ejt$, there are two critical values of doping, $x_{c1}$ and
$x_{c2}$, as shown in \fig{btype}. For $x < x_{c1}$, {\it there are
no occupied $b$-electron states in the system}, and the holes form a
Coulomb glass\cite{CoulombGlass}. For doping larger than $x_{c1}$,
some occupied $b$-electron puddles appear (\Fig{clump}(a)). On
further increase of doping, clumps interconnect and percolate
through the simulation box (\Fig{clump} (b)), giving rise to the
possibility of occupancy of $b$-states extended throughout the
system, and hence metallicity. The results for $x_{c1}$ and $x_{c2}$
are insensitive to $V_0$ for $V_0 \le 0.5$ (the typical value of
$0.01 \le V_0 \le 0.1$ in manganites falls in this range). For
larger values of $V_0$, the clumps are always too small, and are
never occupied by electrons, and the system is an insulator for all
$x < 0.5$ .

\begin{figure}
\centerline{\epsfxsize=6.5truecm \epsfbox{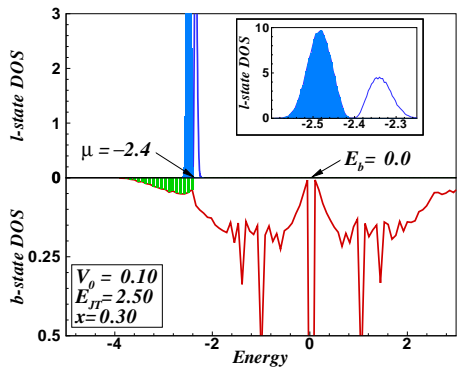} } \caption{(color online) 
Density of states (DOS) of $\ell$ polarons and $b$ electrons
(average over 100 random initial conditions, size: 10$^3$). The occupied
states are shaded. The chemical potential $\mu$ and the $b$ band
center $E_b$ are marked. Inset: soft Coulomb gap of the
polarons.  } \label{dos}
\end{figure}

The physics behind these results is uncovered by a study of the
density of states(DOS) of the $\ell$-polarons and $b$-electrons
shown in \Fig{dos}. We find that the chemical potential $\mu$
%%S of our system
(\Fig{dos}) goes essentially as $\mu = -\Ejt +
V_0$. The polarons form a Coulomb glass which has a soft gap
(\Fig{dos}, inset) at the chemical potential, as in the
classical Coulomb glass\cite{CoulombGlass}. Furthermore, the
polarons have their energies spread over an energy scale of order
$V_0$.\footnote{This
%%S is the reason
why the polaronic states do not lead to any (heavy fermion like)
signatures in the specific heat. Their entropy can be regained only
on thermal energy ($k_B T$) scales of the order of their energy
spread.} The $b$ electrons have a band-like DOS with a bandwidth
that depends on the doping $x$. In fact, we find that the effective
half bandwidth of the $b$ electrons scales as $D_{eff} = D_0
\sqrt{x}$ ($D_0$, the bare half-bandwidth, $= 6t$ for the cubic
lattice), in remarkable agreement with an earlier\cite{Pai2003}
single site dynamical mean field theory(DMFT) result obtained for
\prn{purelb}! Furthermore, the simulation results show that the $b$
band center (marked by $E_b$ in the figure) is {\em not} affected by
doping, and scales as $E_b \approx V_0$ (for small $V_0 \le 0.5$).

The behavior of the chemical potential and the band center are due
to the fact that the random distribution of Ak ions plays an
important role in determining the real space structure. From a study
of the positional correlation function between two holes, and a hole
and an Ak ion, we find that the holes tend to cluster around the Ak
ions\cite{details}. Furthermore, the electrostatic screening is
quite strong in this system in that these correlation functions
(both hole-hole and hole-Ak ion) reach a plateau within a few
lattice spacings. Since the holes cluster around the (oppositely
charged) Ak ions, it is clear that a typical $\ell$ polaron will
have a larger number of $\ell$-polaron neighbors than hole sites.
Thus, the $\ell$ polaron sites see, on an average, an electrostatic
potential of order $V_0$, and the average energy of an $\ell$
polaron is increased from $-\Ejt$ to $-\Ejt + V_0$. Since the
chemical potential of a Coulomb glass is the average energy of the
states\cite{CoulombGlass}, it follows that the chemical potential is
$-\Ejt + V_0$. Similarly every hole site sees an electrostatic
potential of $V_0$, implying that the $b$ band center is placed at
$V_0$. These observations and the fact that the $b$ bandwidth
scales as $\sqrt{x}$, suggests that the minimum doping for the the
$b$-band bottom to touch the chemical potential (i.e., for $b$-state
occupancy) is $x_{c1} = \left(\frac{\Ejt}{D_0} \right)^2$, exactly
the DMFT prediction\cite{Pai2003} for \prn{purelb}!

However, we note that the simulation results include important new
physics not contained in the DMFT\cite{Pai2003}, namely that $b$
state occupancy does {\it not} by itself make the system a metal, as
these states are localized inside the clumps. Based on inverse
participation ratio, geometric percolation of the clumps and Kubo
conductivity calculations\cite{details}, we have good estimates for
a second, higher, critical doping level, $x_{c2}$, at which the
system actually becomes a metal (\Fig{btype}), for which {\it the
occupied b-states should extend across the simulation box}. Our
results suggest that for $x < x_{c1}$, electrical transport should
be that of a classical Coulomb glass ($\sigma \sim
e^{-A/\sqrt{T}}$). For $x_{c1} < x < x_{c2}$, the transport should
have two contributions - the first a Coulomb glass contribution of
the $\ell$ polarons and the second a variable range type
inter-puddle hopping of the electrons in the $b$ puddles, as is
indeed observed in doped manganites\cite{Sudheendra2003}. For $x >
x_{c2}$ we find a highly resistive metal.

Another interesting aspect that we have investigated\cite{details}
is the clump size $R$, and its dependence on  $V_0$. {\it For a
(fictitious) uniform distribution of Ak ions}, one can show by an
approximate analytical calculation that
$R~\sim~\frac{1}{\sqrt{V_0}}$, corresponding, for $\Ejt=2.5$ and
$x=0.3$, to clump sizes  between 10 and 5 lattice spacings for $V_0$
between $0.01$ and $0.1$. The clump sizes for the more realistic,
random, distribution of Ak ions obtained from our simulations are
much smaller; even a very small $V_0$ produces clumps that are four
to five lattice spacings, and this spacing is essentially
independent of $V_0$ for realistic values of $V_0$ - in stark
contrast to the analytical result above. Thus the long range Coulomb
interaction is a `singular perturbation' that prevents macroscopic
phase separation; but the sizes and the distribution of the clumps
are determined by the random distribution of the Ak ions. Thus we
conclude that doped manganites as modeled by \prn{ElbHam} are
necessarily and intrinsically electronically inhomogeneous, on a
{\it nanometric scale}.

Our results have been obtained for a Hamiltonian and energy
parameters that are very realistic especially in low bandwidth
manganites with a large ferromagnetic region in their phase diagram.
In sharp contrast to some of the earlier scenarios proposed in
manganites, the nanoscale inhomogeneities we obtain are {\it not}
due to `phase competition' induced `phase separation' between
`insulating' and `metallic' phases frustrated by disorder as
suggested by studies on `toy models', such as spin
Hamiltonians\cite{Dagotto2003} or Hamiltonians with two localized
states and electron lattice coupling\cite{Ahn2004}. Rather, they
arise due to the the long ranged Coulomb interactions frustrating
the phase separation induced by strong local correlations. The {\it
mechanism}  itself has been discussed in a variety of
contexts\cite{Emery,Dagotto2003}, but to the best of our knowledge,
{\it ours is the first quantitative study on a  realistic model
(that includes dopant ions) for any correlated oxide}.

We emphasize that the nanoscale inhomogeneities we obtain are
present in {\it both} the insulating and metallic phases of doped
manganites between which one has a transition as a function of
doping, at $x_{c2}$; and furthermore, as is clear from
Figs.~\ref{clump} and \ref{btype}, each of these constitutes a
valid, single, thermodynamic phase that is homogeneous on
meso-scales. These results are in conformity with the electron
holography results of Loudon et al.\cite{Loudon2002}, where even the
ferro-metallic state is seen to have interspersed in it nanoscale
`insulating regions', which in the context of the present simulation
are just the $\ell$-clusters\footnote{At half doping, where these
experiments are actually done, the relevant $\ell$ clusters have
orbital/charge order. A modified Hamiltonian that accounts for these
is discussed in O. Cepas {\it et al.}, Phys.~Rev.~Lett.~{\bf 94},
247207 (2005) and Phys.~Rev.~B {\bf 73}, 035218 (2006). But the
basics of the nanoscopic inhomogeneities discussed here would remain
the same.}.

Furthermore, our work suggests that {\it mesoscale phase
separation\cite{Dagotto2003}, or proximity to
multicriticality\cite{murakami}, are not essential for explaining
CMR in manganites}. Given the correspondence between the simulation
results for the extended $\ell b$ model and the (homogeneous) DMFT
results for the simple $\ell b$ model, {\it the single metallic
phase obtained here will show a ferro-metal to para-insulator
transition as well as CMR} due to strong Hund's coupling between the
Mn $t_{2g}$ core spins and the $b$ electrons and thermal
fluctuations. As shown elsewhere\cite{Ramakrishnan2004}, the CMR
arises from small field induced changes in the energetics of the
extended $b$ states which are responsible for charge transport, and
the consequent exponential changes in their occupancy. Indeed, by
way of experimental confirmation, we note that there are many
manganites without mesoscale inhomogeneities that show colossal
responses\cite{Mathieu2004}. Finally, our work reinforces
suggestions that the mesoscale patterns seen in experiments arise
from other sources such as long range elastic strains, possibly due
to defects\cite{Strain,Sarma2004}. In our model, they can be
generated from the strain dependence of the local energy parameters
such as $\Ejt$ and $t$. Investigations along these lines are in
progress.

We thank the INSA (VBS), DST and IFCPAR (HRK) and DAE (TVR) for
support.


\begin{thebibliography}{99}
\bibitem{details} The details will be published elsewhere.
\end{thebibliography}
\end{document}